%% file: main.tex
\documentclass[pmlr,twocolumn,10pt]{jmlr} % W&CP article
\usepackage{booktabs}
\usepackage{longtable}
\usepackage{siunitx}

\usepackage[switch]{lineno}

\theorembodyfont{\upshape}
\theoremheaderfont{\scshape}
\theorempostheader{:}
\theoremsep{\newline}

\title[Linking Negative Language to Diagnostic Disparities]{Bias Detection in Emergency Psychiatry: Linking Negative Language to Diagnostic Disparities}

\author{%
\Name{Alissa Valentine} \Email{alissa.valentine@di.ku.dk}\\
\addr Copenhagen University, Denmark and Mount Sinai School of Medicine, USA
\AND
 \Name{Lauren Lepow} \\
 \Name{Donald Apakama} \\
 \Name{Lili Chan} \\
 \Name{Alexander Charney} \\
 \Name{Isotta Landi} \\
 \addr Mount Sinai School of Medicine, USA
}

\begin{document}

\maketitle

\begin{abstract}
The emergency department (ED) is a high stress environment with increased risk of clinician bias exposure. In the United States, Black patients are more likely than other racial/ethnic groups to obtain their first schizophrenia (SCZ) diagnosis in the ED, a highly stigmatizing disorder. Therefore, understanding the link between clinician bias exposure and psychiatric outcomes is critical for promoting nondiscriminatory decision-making in the ED. This study examines the association between clinician bias exposure and psychiatric diagnosis using a sample of patients with anxiety, bipolar, depression, trauma, and SCZ diagnoses (N=29,005) from a diverse, large medical center. Clinician bias exposure was quantified as the ratio of negative to total number of sentences in psychiatric notes, labeled using a large language model (Mistral). We utilized logistic regression to predict SCZ diagnosis when controlling for patient demographics, risk factors, and negative sentence ratio (NSR). A high NSR significantly increased one’s odds of obtaining a SCZ diagnosis and attenuated the effects of patient race. Black male patients with high NSR had the highest odds of being diagnosed with SCZ. Our findings suggest sentiment-based metrics can operationalize clinician bias exposure with real world data and reveal disparities beyond race or ethnicity.
\end{abstract}

\paragraph*{Data Availability}
Our study utilizes EHR data from a large, diverse medical center. Details on the data and preprocessing steps are provided in the following sections. The data cannot be shared due to data use agreement.

\paragraph*{Institutional Review Board (IRB)}
This project was reviewed and approved by the IRB. All methods were performed in accordance with the relevant IRB guidelines. This project was reviewed and approved by the Mount Sinai IRB via STUDY-20-00338. All methods were performed in accordance with the relevant guidelines. Informed consent was waived by the IRB. 

\section{Introduction}
\thispagestyle{empty}
\label{sec:intro}
The emergency department (ED) is a high stress environment where clinicians are expected to make quick decisions. The ED environment especially exacerbates physicians’ cognitive functioning, increasing their implicit racial bias, and perpetuating racial disparities during decision-making \citep{RN286}. For instance, in the United States, Black patients are more likely to receive their first schizophrenia diagnosis in the emergency setting compared to other racial and ethnic groups \citep{RN222,RN23,RN235,RN348}.

Schizophrenia spectrum disorders (SCZ) are highly stigmatizing \citep{RN4,RN102}, and decades of research has demonstrated notable racial disparities in their diagnostic rates. In the United States, the SCZ diagnosis rate is 2-4 times greater in Black patients than in White patients \citep{RN179,RN6,RN15,RN71,RN345}. Explanations for the disparities in SCZ diagnosis rates often emphasize the role of clinician bias on diagnostic decision-making. Here we define clinician bias as an implicit or explicit belief held by a clinician about a patient based on the patient’s sociodemographic characteristics that prevents the clinician from impartial clinical decision-making during the diagnosis process \citep{RN30}. 

Some attribute SCZ’s diagnostic disparities to misdiagnosis, wherein the patient could have otherwise obtained a mood disorder diagnosis like depression \citep{RN175,RN171,RN201,RN213,RN87,RN200}, or a stress or dissociative disorder due to a history of trauma \citep{RN360,RN361,RN358,RN359,RN83,RN363}. This could be driven by clinicians recognizing symptoms differently in Black versus White patients \citep{RN218, RN221,RN220,RN213,RN173}, suggesting that clinician bias influences the diagnostic process of SCZ. As such, there is a need to address how exposure to clinician bias is related to psychiatric outcomes in the ED to ensure nondiscriminatory diagnosis.

Psychiatric notes capture the signs, symptoms, and behaviors of patients from the perspective of the clinician due to the observational nature of psychiatric decision-making. When documenting the clinical encounter, the language used by clinicians can be classified as neutral, negative, or positive \citep{RN75}. Advances in natural language processing (NLP) methods have led to more efforts to quantify biased or harmful language use in clinical text. Recent work has shown that Black patients obtaining care in the ED are more likely to be described with negative or stigmatizing words in clinical notes compared to other racial or ethnic groups \citep{RN16,RN296}, essentially capturing patient exposure to clinician bias. However, to our knowledge no studies have leveraged NLP methods to explore biased language use in real world psychiatric data. Doing so is critical to understand the extent to which clinician bias is embedded in the clinical notes of psychiatric SCZ patients.

In this study, we examine patient exposure to clinician bias in the ED using electronic health record (EHR) data from a large, diverse health care system. We leverage a large language model (LLM) to label the sentiment of sentences describing a patient in their first ED Psychiatric Note written by a clinician, quantifying clinician bias exposure as the ratio of negative to total number of sentences describing a patient, or negative sentence ratio (NSR). We then used logistic regressions to investigate the relationship between the NSR and SCZ diagnosis in a cohort of patients diagnosed with anxiety, bipolar, depression, trauma, and SCZ in the same ED setting. When doing so, we took into consideration known risk factors for SCZ. Furthermore, we implemented an intersectional framework to investigate how compounding demographic variables drive diagnostic disparities in ED psychiatry. The contributions of this work:
\begin{enumerate}
    \item Reveal that the racial disparities in SCZ diagnosis are moderated by metrics of clinician bias exposure in the ED.
    \item Assess the association between negative language use and psychiatric diagnoses in the ED, showing that SCZ diagnoses have the strongest association with negative patient descriptions and trauma diagnoses have the weakest.
    \item Demonstrate that sentiment-based metrics can detect exposure to clinician bias in real world EHR data.
\end{enumerate}

\section{Related Work}
\subsection{SCZ Risk Factors}
Previous work has demonstrated that age, sex, race, socioeconomic status (SES), and history of trauma or substance use contribute to SCZ rates in diverse medical settings in the United States \citep{RN345,RN6}. Such work highlights the need to invoke an intersectional lens when investigating SCZ diagnosis rates. Intersectionality is a framework for recognizing that one's lived experiences are influenced by the combination of one's many identities (i.e., race, sex, gender, ability, religion, language, etc.), often reflecting one's experiences of privilege and discrimination \citep{crenshaw2013}. For instance, \cite{RN345} found that males have a higher risk of obtaining a SCZ diagnosis than females, as well as Black patients have a higher risk than White patients. However, their results demonstrated that patient sex, race, and SES also interact in the prediction of SCZ, such that high SES acted as a protective buffer against SCZ diagnosis for White patients but not Black patients. This contradicts previous work suggesting that high SES decreases one's risk of obtaining a severe psychiatric diagnosis \citep{RN91}. 
A history of trauma or substance use is highly co-morbid with SCZ, and generally considered a risk factor for developing SCZ as well \citep{RN37,RN42,RN83,RN84}. Furthermore, \cite{RN84} demonstrated an interaction between trauma and substance use wherein they argued a "double hit" of both may increase risk of developing SCZ.
Lastly, existing work from the All of Us Research Program has shown that the odds of obtaining a SCZ diagnosis decreases as you get older \citep{RN6}, reflecting that the peak age of SCZ symptom onset is between 20-29 years of age \citep{miettunen2018age,RN81}. 

\subsection{Clinician Bias Detection}
Whilst clinician bias cannot be directly measured, proxy variables can capture patterns that are linked to bias exposure, offering a practical way to capture its influence on clinical outcomes. Clinical notes offer an opportunity to quantify clinician bias exposure, as they are written by the clinician and reflect their perspective of the patient. The language the clinician uses to describe the patient can be classified as neutral, negative, or positive \citep{RN75}. Negative patient descriptors include those that question patient credibility, reasoning, insight, or judgment; portray the patient as noncompliant or as a threat; remark on the patient’s poor self-care; or generally conveys disapproving feelings towards the patient and their presentation. In contrast, positive patient descriptors include patient strengths, minimization of blame, and language that conveys of approval and positive feelings towards the patient and their presentation. Recent work by \cite{APAKAMA2025100296} has shown that LLMS offer scalable methods of detecting language that discredits, stigmatizes, and stereotypes patients in the ED. However, such methods have yet to be applied to the psychiatric domain.

\subsection{Sentiment Analysis}
Sentiment analysis is a robust approach to quantifying the tone conveyed in text as positive, neutral, or negative. Outside of medicine, sentiment analysis is often deployed to mine the attitudes of people on social media and identify harmful or discriminatory text \citep{RN88}. Within psychiatry, \cite{RN352} adapted sentiment analysis with deep learning models to quantify clinician’s attitudes towards a patient’s prognosis across domains such as appearance, thought content, substance use, and more. They found that including metrics of clinical sentiment in machine learning models improved their performance when predicting hospital readmission in psychiatric patients with psychosis \citep{RN343}. Our recent work has shown that large language models (LLMs) outperform pretrained language models (PLMs) on sentiment analysis tasks with clinical text describing psychiatric patients \citep{RN344}.
% to be cited later: \citep{RN344}.

\section{Study Design}
\subsection{Data}
Structured and unstructured data was queried from the data warehouse of a large, diverse medical center and comprised of data from January 2012 to October 2023. The dataset includes structured patient and encounter data: patient sex, race, ethnicity, date of birth, zip code at encounter, date of encounter, primary ICD-10 diagnosis, and note type. To control for differences in note author type, content, and structure, we chose to only use clinical notes with a note type of “ED Psychiatric Note". There is typically only one "ED Psychiatric Note" per ED encounter written by a psychiatrist if the patient presents to the ED with psychiatric care needs. Therefore, this note type contains the information about the patient most relevant to their psychiatric care. As such, we chose to use this note type, contrary to "Progress Note" or "Discharge Note" types.
%Mount Sinai’s Caboodle Data Warehouse and comprised of data from January 2012 to October 2023. The dataset includes structured patient and encounter data: patient sex, race, ethnicity, date of birth, zip code at encounter, date of encounter, primary ICD-10 diagnosis, and note type. To control for differences in note structure and content, we chose to only use clinical notes with “ED Psych Note” in “Note Type.” Notes with less than 500 words and over 3000 words were dropped. To account for data entry errors, patients were dropped if their age at the time of the ED note was less than 1 or greater than 100.

\subsubsection{Cohorts}
Patients were included in the SCZ cohort if the primary diagnosis code in the encounter that contained their first ED Psychiatric Note belonged to the “Schizophrenia spectrum and other psychotic disorders” category of the Clinical Classifications Software Refined (CCSR; \cite{ccsr}). The CCSR is a tool which groups ICD-10 diagnosis codes into clinically meaningful categories. By using the CCSR, we can study schizophrenia spectrum disorders rather than schizophrenia alone. 

Patients were included in the control cohort if they never obtained a SCZ diagnosis and if the primary diagnosis code in the encounter that contained their first ED Psychiatric Note belonged to the “Anxiety and fear-related disorders”, “Bipolar and related disorders”, “Depressive disorders”, and “Trauma- and stressor-related disorders” categories of the CCSR. These diagnostic categories were chosen due to their symptom overlap with SCZ and to reflect previous hypotheses on misdiagnosis of SCZ \citep{RN360,RN350,RN349,RN213}. 

\subsubsection{Covariates}
A history of trauma and substance use is attributed only to patients with a trauma-related or substance diagnosis before their first ED Psychiatric Note. Trauma diagnoses are defined as the ICD-10 codes mapped to CCSR category “Trauma- and stressor-related disorders”. Substance use diagnoses are the ICD-10 codes mapped to CCSR categories “Alcohol-related disorders”, “Opioid-related disorders”, “Cannabis-related disorders”, “Sedative-related disorders”, “Stimulant-related disorders,” “Hallucinogen-related disorders”, “Tobacco-related disorders”, “Inhalant-related disorders”, and “Other specified substance-related disorders.”

SES is defined using patient zip code, which is mapped to the United States Census Bureau for median-household income measures in 2021 that range from \$21,846-\$250,000 \citep{census}. The median household-income is a valid measure of patient SES and is related to health outcome disparities \citep{RN11}.

Our EHR data provided a race and an ethnicity for each patient. Race typically included values denoting race or nationality (e.x., "African American", "Asian", "Black", "Chinese", "White", etc.). Ethnicity included values denoting Hispanic identity (i.e., "Hispanic/Latino", "Non-Hispanic"). We chose to merge the race and ethnicity data to create one Race/Ethnicity value per patient. The methods of merging and categorizing race and ethnicity data were based off Office of Management and Budget Statistical Policy Directive No. 15: Standards for Maintaining, Collecting, and Presenting Federal Data on Race and Ethnicity \citep{OMB}. Due to small sample sizes, the final dataset combined multiple categories into Hispanic/Latino (Hispanic/Latino, White Hispanic, Black Hispanic, American Indian or Alaska Native Hispanic, and Native Hawaiian or Pacific Islander Hispanic) and Some Other Race (Some Other Race, Native Hawaiian or Pacific Islander, American Indian or Alaska Native, and Middle Eastern or North African). The final Race/Ethnicity groups and their sample sizes can be seen in Table \ref{tab:demo}, which shows the demographic make up of the SCZ cohort and the control cohort consisting of non-SCZ psychiatric (psych) patients.

\subsection{Text Preprocessing}
Clinical notes in both the SCZ and control dataset were preprocessed into sentences for future labeling. The first step in this process was to adapt the medspaCy sectionizer \citep{RN341} to extract the following sections from each note, if available: Mental Status Exam (MSE), History of Present Illness (HPI), Chief Complaint (CC), and Collateral. These sections were selected because they most often contain free text written by the physician that describes the patient and their conversations. The Collateral section was included because it contains free text written by the physician that describes information about the patient shared by someone other than the patient, which takes place if the patient is incapacitated or unable to provide their medical history. In the psychiatric ED, collateral is also more commonly collected to enable clinical decision making is not purely based on seeing the patient for a few hours but incorporates patient history from someone who, ideally, knows them well. The extracted free text sections were parsed by the medspaCy sentencizer \citep{RN341}, outputting the free text paragraphs into sentences. The sentence corpus was input to a pipeline that classifies sentences as clinically relevant or irrelevant, removing the clinically irrelevant sentences from the main corpus \citep{RN340}. Upon exploring the sentences labeled clinically irrelevant, we found most of them contained contact or scheduling information, and sentences from the MSE section that were not descriptive of the patient. Following sentence preprocessing, patients were dropped if their note had less than 4 sentences.
% to be cited \citep{RN340}

\subsection{Sentiment Analysis}
\subsubsection{Model Selection}
Mistral-7B-Instruct-v0.2 was used in distributed inference to label the sentiment of the final sentence corpus with a prompt-based approach. Mistral is an open-source model and was accessed via Hugging Face. The temperature of the model was set to 0.001 and seed to 42 to enhance reproducibility. Mistral was selected due to our previous work which demonstrates that it out performs other LLMs on sentiment analysis tasks with psychiatric patient descriptions \citep{RN344}. In such work, five LLMs were compared in terms of their performance labeling the sentiment of psychiatric patient descriptions. Two labels were used in that dataset, one of physicians and another of non-physicians, thus exploring how some models align more towards the physician or non-physician point of view of the sentiment of psychiatric text. Mistral was also found to align best to the non-physician point of view of the sentiment of psychiatric patient descriptions.
% to cite: \citet{RN344}

\subsubsection{LLM Prompts}
We used the prompt from our previous work that asks the LLM to assume the identity of a patient reading their own clinical note \citep{RN344}. By deploying the model to label sentiment from the perspective of patients, we argue it brings us closer to equitable bias quantification in the psychiatric domain. The prompt is seen below:

\textbf{User:} \textit{"As a patient at a medical center, medical doctors write lots of clinical notes about you. Your task is to analyze the sentiment of a series of sentences your doctor wrote about you. For each sentence, how do you feel reading this description of you? Please assign a sentiment score of negative, neutral, or positive for each sentence."}

\subsubsection{Parsing LLM Output}
Model outputs were parsed using JSON formatting, with $<$1\% of sentences not obtaining a sentiment label due to model behavior. These sentences were labeled “NA”. Mistral’s NA output was found to have five categories:
\begin{enumerate}
    \item TWO LABELS: The model output more than one label (\textit{e.x., \{"0":"neutral-negative"\}}).
    \item ERROR: The model did not output a sentiment label (\textit{e.x., \{"sentence": "pt presents with..."\}}).
    \item EXTRACT: The model output a sentiment label with incorrect JSON format (\textit{e.x., \{"0":"neutral, explanation: ..."\}}).
    \item REFUSE: The model refused to label the sentence due to it containing explicit content (\textit{e.x., \{"I cannot analyze a sentence that contains information about a patient's sexual assault."\}}).
    \item EXTRANEOUS TEXT ($E_{TEXT}$): The model expressed inability to label the sentence due to it being incomplete, containing uninterpretable medical jargon, or being made of text irrelevant for sentiment labeling (\textit{e.x., \{"I'm unable to determine the sentiment from an incomplete sentence."\}}).
\end{enumerate}
Approximately 60\% of NA sentences were identified as $E_{TEXT}$. For sentences with the “EXTRACT” subtype, we used a rule-based approach to look for “negative,” “neutral,” or “positive” strings in the text. Mistral was rerun on sentences with “TWO LABELS”, “ERROR”, and “REFUSE” subtypes using a prompt that emphasized returning only one sentiment score with specific JSON formatting. After the end of these steps, any sentences with a remaining NA label were kept for further analysis. Sentences with NA output subtype “$E_{TEXT}$” were thrown out from the dataset, due to the sentences consisting of text that was irrelevant for a sentiment analysis task such as text purely used for structuring the contents of the clinical note, or listing contact information for patient referrals.

\subsubsection{Manual Validation of LLM Labels} \label{sec:Manual Validation of LLM Labels}
After obtaining the sentiment labels from Mistral, we performed a round of manual validation on 30 sentences for each label (negative, neutral, positive). Two authors, one physician and one non-physician, performed the manual validation. Mistral performs better than chance on all three labels as seen in Table \ref{tab:llm_valid}. Interestingly, Mistral displays high precision on positive and negative sentences, but low precision on neutral. The model also has high recall on neutral and negative sentences but low recall on positive sentences. This suggests the model often mislabels neutral sentences as positive and negative, and mislabels positive sentences as neutral. Based on these results, the rest of the analyses only utilized the negative sentiment labels from Mistral.

\input{model_performance}

\subsection{Bias Metric}
Following the sentiment analysis manual validation in Section \ref{sec:Manual Validation of LLM Labels}, the negative class label was found to have the highest precision and recall, with a significant number of false positives and false negatives on neutral and positive labels. To address this, the investigators decided to create a bias exposure metric that prioritized the amount of negative labeled sentences from Mistral’s output and the number of total sentences per patient note. The result was the Negative Sentence Ratio (NSR), calculated as seen below:
\begin{equation}
n_{sentences} = n_{NA} + n_{negative} + n_{neutral} + n_{positive}
\label{eq:one}
\end{equation}

\begin{equation}
NSR = \frac{n_{negative}}{n_{sentences}}
\label{eq:two}
\end{equation}

\subsection{Association Analysis}
In this project, we hypothesized that including a metric for clinician bias exposure would reduce the association between patient race/ethnicity and SCZ diagnosis. We also suspected patient race/ethnicity, sex, and SES to have significant interactions with one’s exposure to clinician bias. Lastly, we considered that a history of trauma and a history of substance use may interact to impact risk of obtaining a SCZ diagnosis.

We performed two logistic regressions to test these hypotheses, one with and one without the NSR variable. In the second model, we removed the NSR and its interaction terms from the algorithm to understand the impact of controlling for clinician bias exposure in predicting SCZ diagnosis and serve as a baseline comparison when interpreting our results. Both models used the White category within Race/Ethnicity and the Female category within Sex as the references. The formulas used by the two models are detailed below.

\textbf{Model 1:}
\begin{align*}
\operatorname{logit}\!\big(P(\text{SCZ})\big) 
            &= \beta_0 
             + \beta_1 \,\text{Age}
             + \beta_2 \,\text{Sex}
             \nonumber\\
             &\quad
             + \beta_3 \,\text{Race/Ethnicity}
             + \beta_4 \,\text{SES}
             \nonumber\\
             &\quad
             + \beta_5 \,\text{Trauma}
             + \beta_6 \,\text{Substance}
             \nonumber\\
             &\quad
             + \beta_7 \,\text{Trauma:Substance}
             \nonumber\\
             &\quad
             + \beta_8 \,\text{Race/Ethnicity:Sex}
             \nonumber\\
             &\quad
             + \beta_9 \,\text{Race/Ethnicity:SES}
             \nonumber\\
             &\quad
             + \beta_{10} \,\text{Sex:SES}
             \nonumber\\
             &\quad
             + \beta_{11} \,\text{Race/Ethnicity:Sex:SES}
             \nonumber\\
             &\quad
             + \beta_{12} \,\text{NSR}
             \nonumber\\
             &\quad
             + \beta_{13} \,\text{NSR:Race/Ethnicity}
             \nonumber\\
             &\quad
             + \beta_{14} \,\text{NSR:Sex}
             \nonumber\\
             &\quad
             + \beta_{15} \,\text{NSR:SES}
\end{align*}

\textbf{Model 2:}
\begin{align*}
\operatorname{logit}\!\big(P(\text{SCZ})\big) 
            &= \beta_0 
             + \beta_1 \,\text{Age}
             + \beta_2 \,\text{Sex}
             \nonumber\\
             &\quad
             + \beta_3 \,\text{Race/Ethnicity}
             + \beta_4 \,\text{SES}
             \nonumber\\
             &\quad
             + \beta_5 \,\text{Trauma}
             + \beta_6 \,\text{Substance}
             \nonumber\\
             &\quad
             + \beta_7 \,\text{Trauma:Substance}
             \nonumber\\
             &\quad
             + \beta_8 \,\text{Race/Ethnicity:Sex}
             \nonumber\\
             &\quad
             + \beta_9 \,\text{Race/Ethnicity:SES}
             \nonumber\\
             &\quad
             + \beta_{10} \,\text{Sex:SES}
             \nonumber\\
             &\quad
             + \beta_{11} \,\text{Race/Ethnicity:Sex:SES}
\end{align*}

In addition, we performed 4 logistic regressions using the NSR as the only predictor for each diagnosis in our cohort (i.e., Anxiety, Bipolar, Depression, Trauma, and SCZ) to compare how exposure to clinician bias in the ED is associated with each diagnosis in the context of all other diagnoses. 

Post-hoc analyses utilized Fisher’s one-way ANOVA to explore differences between Race/Ethnicity groups for the NSR, number of sentences, number of negative sentences, and number of NA sentences per patient. Odds ratios (ORs) and 95\% confidence intervals (CIs) are calculated using the exponents of the coefficients. When reporting the effect of interactions, we use the exponent of the coefficient of the interaction term.

Lastly, notes with less than 500 words and over 3000 words were dropped to control for notes with abnormal length, and to reflect the normal curve seen in our distribution. To account for data entry errors, patients were dropped if their age at the time of the ED note was less than 1 or greater than 100.

\input{demo_table} \label{tab:demo}

\begin{figure*}[htbp]
\floatconts
  {fig:nodes}
  {\caption{Marginal Effects of Patient Race/Ethnicity and Sex on SCZ Diagnosis at levels of NSR.}}
  {\includegraphics[width=\linewidth]{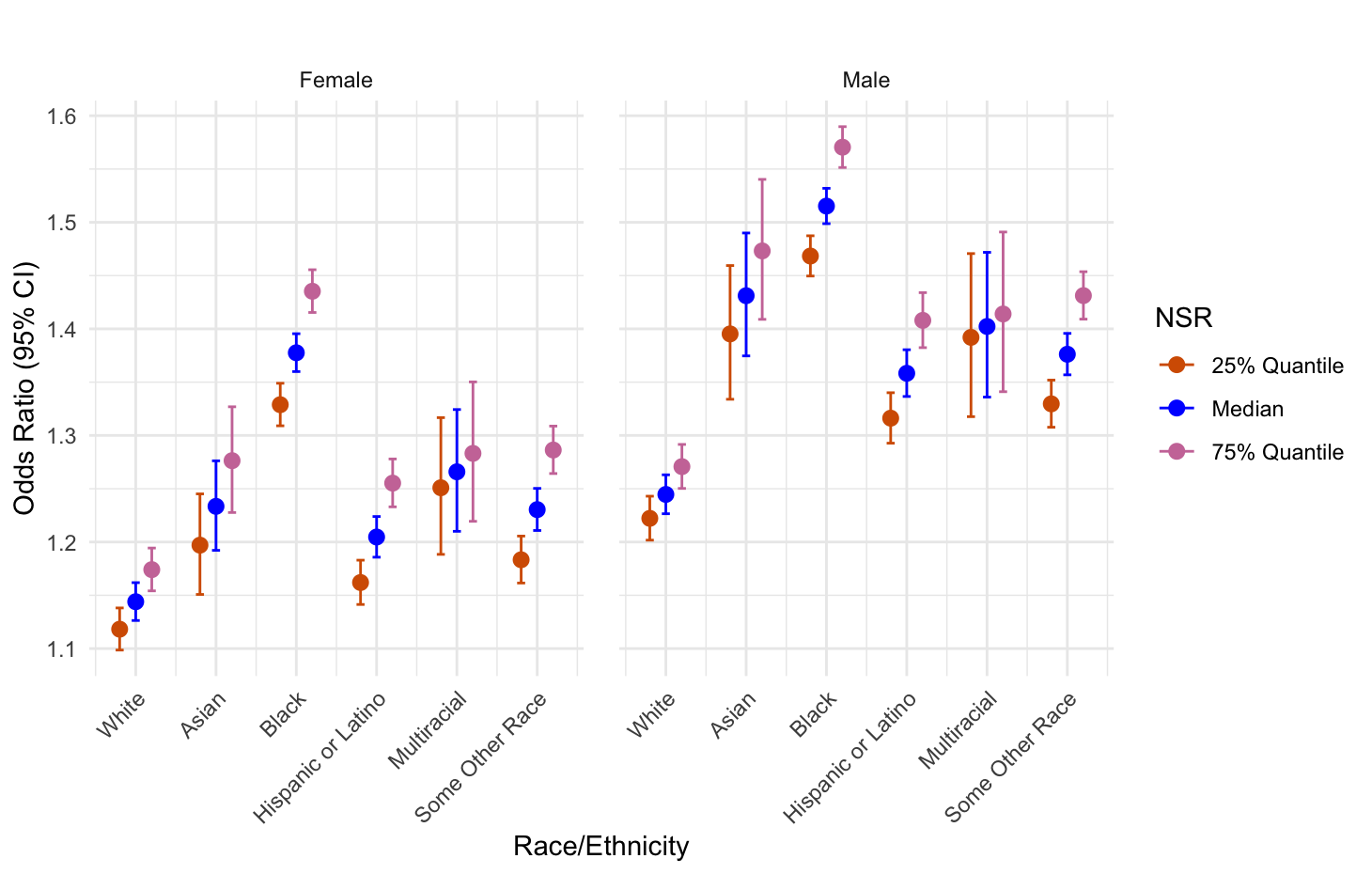}}
  \label{fig:race_nsr_scz}
\end{figure*}

\section{Results}

The sample population consisted of 29,005 patients with a primary diagnosis of Anxiety, Bipolar, Depression, Trauma, or SCZ disorders in their first ED Psychiatric Note. There were 8,114 patients in the SCZ cohort, of which most identified as Black (41.2\%) or male (60.2\%). The control cohort was also majority Black (28.7\%) and almost equally split between males (50.1\%) and females (49.9\%). SCZ patients were significantly older (40.67[SD=15.39]) than the controls on average ($p<0.001$). SCZ patients carried more negative sentences (7.76 [SD=4.56]) on average ($p<0.001$). This led to a significantly higher average NSR for SCZ patients (0.28 [SD=0.14]) compared to patients without SCZ diagnosis ($p<0.001$). See Table \ref{tab:demo} for more.

\subsection{Model Comparison}
The prediction model including the NSR and its interaction terms was significant (\(AIC=29429,R^2=0.20,F(35,28969)=207.6,p<0.001\)), and demonstrated improved performance compared to the model without the NSR and its interaction terms (\(AIC=29808,R^2=0.19,F(27,28977)=251,p<0.001\)). In the following results, the ORs and CIs are reported using coefficients from the model including the NSR terms if the finding was the same in both models. For full model comparison, see Table \ref{tab:regression} in Appendix.

\subsection{NSR}
When controlling for other variables, the NSR was the strongest predictive term for SCZ diagnosis in the model (\(OR=1.34 [95\% CI=1.52-1.19 ],p<0.001\)). The NSR significantly interacted with patient race/ethnicity such that a high NSR increases the odds of SCZ diagnosis greater in those identifying as Black, Hispanic or Latino, and Some Other Race. See Figure \ref{fig:race_nsr_scz} for more. There were no significant interactions between the NSR and other variables.

\subsection{Sociodemographic Factors}
Including the NSR in the model changed the role of patient Race/Ethnicity in predicting SCZ diagnosis. More specifically, in the SCZ prediction model without the NSR, Black patients are significantly more likely to obtain a SCZ diagnosis (\(1.10 [1.15-1.05],p<0.001\)). However, this association is no longer significant when including the NSR in the model. See Table \ref{tab:regression} in the Appendix for further comparison.

There was a significant interaction between sex and Race/Ethnicity such that patients identifying as Black \textit{and} male have increased odds of obtaining a SCZ diagnosis in the ED (\(OR=1.08 [1.15-1.01],p<0.05\)). Identifying as Asian \textit{and} male also increased one’s odds of SCZ diagnosis (\(OR=1.20 [1.38-1.04],p<0.05\)). However, due to a small sample of Asian-identifying patients in our cohort (n=975), and large confidence interval, it is difficult to interpret this finding as clinically relevant, and we call on further research to expand on this.

Lastly, older age was associated with significantly higher odds of receiving a SCZ diagnosis. For example, individuals an age of 24 (the 25\% quartile of age) had an OR of 1.06 (1.07-1.06), compared to 1.14 (1.15-01.12) for those with an age of 49 (the 75\% quartile of age). Male patients had higher odds of obtaining SCZ diagnosis than females (\(OR=1.14 [1.20-1.08],p<0.001\)). 

\subsection{Socioeconomic Status}
SES, measured as the median household income associated with one’s zip code, was not found to have a significant association with SCZ diagnosis on its own. However, the interaction of SES and Race/Ethnicity had a significant effect on odds of SCZ diagnosis for those identifying as Black, Hispanic/Latino, and Some Other Race. Lastly, there were two significant three-way interactions such that male, Asian or Black, high SES patients had decreased odds of obtaining a SCZ diagnosis. See Figure \ref{fig:ses_sex_race} in Appendix, wherein we find that high SES does not act as protective buffer for Black, female patients against obtaining a SCZ diagnosis.

\subsection{History of Trauma or Substance Use}
In both models, having a previously documented diagnosis of trauma-related disorder or substance use disorder significantly decreased one’s odds of obtaining a SCZ diagnosis in the ED. However, the interaction term showed that patients with a history of both diagnoses had higher odds of SCZ diagnosis (\(1.11 [1.14-1.08],p<0.001\)). See Table \ref{tab:regression} in the Appendix for more.

\subsection{NSR Association with Anxiety, Bipolar, Depression, Trauma, and SCZ}
When using the NSR as the only predictor, the prediction models for anxiety, bipolar, depression, trauma, and SCZ diagnoses were significant (Anxiety: \(R^2=0.02,F(1,29003)=516.8,p<0.001\); Bipolar: \(R^2=0.004,F(1,29003)=122.5,p<0.001\); Depression: \(R^2=0.01,F(1,29003)=335.6,p<0.001\); Trauma: \(R^2=0.03,F(1,29003)=1012,p<0.001\); SCZ: \(R^2=0.03,F(1,29003)=954.9,p<0.001\)). As seen in Table \ref{tab:diagnosis-odds}, our results demonstrate that an increased NSR significantly decreases one’s odds of obtaining an anxiety or trauma diagnosis, but increases one’s odds of obtaining a bipolar, depression, or SCZ diagnosis when not controlling for patient sociodemographic group, SES, or other risk factors.

\input{diagnoses_or_table}

\subsection{Group Comparisons}
Our post-hoc analyses explored group differences in the average number of sentences, number of negative sentences, number of NA sentences, and NSR. A significant difference was found between the average number of sentences per patient between groups (\(F_{Fisher} (5,28999)=46.22,p<0.001,\widehat{\omega_p^2}=7.73×10^{-3}\)). We found that patients identifying as Asian had significantly more sentences on average (30.1 [SD=9.3]) compared to other groups. In contrast, Black patients had significantly fewer number of sentences on average (27.1 [SD=9.1]). Similarly, a significant difference was found between the average number of negative sentences per patient between groups (\(F_{Fisher} (5,28999)=14.48,p<0.001,\widehat{\omega_p^2}=1.36×10^{-3}\)). Asian patients (7.1 [SD=4.5]) and those in the Some Other Race group (7.0 [SD=4.4]) had more negative sentences on average, meanwhile Black (6.6 [SD=4.3]) and Hispanic/Latino patients (6.5 [SD=4.3]) had the fewest negative sentences on average. Altogether, this led to a significant difference between the average NSR per patient between groups (\(F_{Fisher} (5,28999)=12.51,p<0.001,\widehat{\omega_p^2}=1.98×10^{-3}\)). Patients with their Race/Ethnicity categorized as Black or Some Other Race had the largest NSR on average (0.24 [SD=0.14]). There were no significant group differences in the number of NA sentences. See Figures \ref{fig:group_n_sent}, \ref{fig:group_neg_sent}, and \ref{fig:group_nsr} in the Appendix for more group comparisons.

\section{Discussion}
In this study, we investigated a cohort of psychiatric patients in the ED with anxiety, bipolar, depression, trauma, and SCZ diagnoses to explore how exposure to clinician bias is associated with obtaining a stigmatizing diagnosis such as SCZ. Our proxy for clinician bias exposure was the negative sentence ratio (NSR). We found that the NSR was the strongest predictor for SCZ diagnosis in the ED when controlling for patient sex, Race/Ethnicity, and known risk factors. Our work demonstrates that detecting patient exposure to clinician bias is not only operational with real world data, but critical to account for in environments where patients are at a high risk of obtaining a stigmatizing diagnosis like SCZ.

Contrary to our expectations and previous work by \citet{RN16} and \citet{RN296}, Black patients did not have more negative sentences in their clinical notes than other racial/ethnic groups (although Black patients did have a \textit{higher proportion} of negative sentences). Instead, patients identifying as Asian and Some Other Race had the most negative sentences. Some Other Race is an aggregation of several distinct racial/ethnic groups in our patient population that we did not want to drop from our experiments due to small sample sizes (i.e., Some Other Race, Native Hawaiian or Pacific Islander, American Indian or Alaska Native, and Middle Eastern or North African). Although it is difficult to interpret the results in the Asian patient population due to small sample size (see Table \ref{tab:demo}), these findings suggest that patients who are more likely to be unaccounted for in research (i.e. "Some Other Race") may carry the highest risk of being exposed to clinician bias. Future work on bias and disparities should prioritize such groups.

Although Black patients did not have most negative sentences compared to other groups, controlling for the NSR attenuated the effect of race on SCZ diagnosis. This finding may be explained by post hoc analyses wherein Black patients were found to have fewer sentences written about them in the free text section of ED Psychiatric Notes compared to other racial/ethnic groups. This led to Black patients having a higher proportion of negative sentences compared to other groups. Whilst the reason for the fewer sentences being written about Black patients remains to be determined, future work might explore the association between quality of care and note length. Altogether, our findings suggest the NSR captures variance related to bias exposure that may otherwise be accounted for by patient race in the model. This brings us closer to understanding the confounding factors related to racial disparities in psychiatry.

This project also provides compelling evidence of the “Diminished Return Theory” in a psychiatric setting. This theory states there is a difference between racial/ethnic driven disparities and SES-driven disparities such that patients in racial or ethnic minority groups don’t benefit from increased SES in the same way as White patients. Examples can be seen in depression and other research fields \citep{RN310,RN309,RN314,RN315}. As seen in Figure \ref{fig:ses_sex_race} in the Appendix, our work suggests that higher SES does not act as a protective buffer against SCZ diagnosis for several intersectional groups, namely Black female patients.

A high NSR was also associated with obtaining a depression or bipolar diagnosis. In contrast, a high NSR had a protective effect on the odds of obtaining an anxiety or trauma-related diagnosis. This is a thought-provoking finding, as many researchers have suggested that some patients are misdiagnosed with SCZ instead of obtaining a less stigmatizing diagnosis like depression or trauma-related disorders \citep{RN175,RN171,RN201,RN213,RN87,RN200,RN360,RN361,RN358,RN359,RN83,RN363}. These findings could be interpreted to reflect the stigmatization associated with each diagnosis. However, we did not control for patient demographics when exploring the associations between the NSR and the other diagnoses besides SCZ. Further investigation is needed to explain how the NSR interacts with patient demographics when predicting other psychiatric diagnoses.

There are several limitations within this study. One could argue that negative sentiment is not a reflection of bias, as patients with SCZ are more likely to be described negatively due to more severe symptoms compared to other diagnoses. However, we found during this experiment that Mistral refused to label the sentiment of sentences with the most explicit patient descriptions (i.e., self-harm, assault, etc.). These sentences took place most often in the HPI or Collateral section of the ED Psychiatric Notes, where the patient’s presentation to the ED is described. More work is needed to study how AI refusal behavior impacts LLM deployment in clinical settings, however it’s possible this led to the omission of sentiment labeled sentences from the final dataset wherein the most severe symptoms would have been discussed. Therefore, the explanation that more severe symptoms in SCZ drove the association between the NSR and SCZ is less likely. Furthermore, some argue that biased symptom recognition in Black patients plays a role in the misdiagnosis in SCZ \citep{RN220,RN173,RN218}. If true, then symptom severity may be associated with clinician bias exposure, and could attenuate the effect of race or ethnicity on risk of SCZ diagnosis. Future work could address this hypothesis by using NLP methods to extract symptom information and include these features in our models.

Lastly, there are many ethical concerns when deploying LLMs for bias quantification. Of these we would like to highlight that LLMs are known to perpetuate societal biases in medical contexts \citep{RN351,RN336,RN339}. With the increasing use of LLMs in healthcare research, we need more discussions and resources dedicated towards assessing how the use of biased language in clinical notes threatens equitable deployment of LLMs in medicine. If we don’t take action, these models risk perpetuating the racial disparities demonstrated in this paper. In this spirit, we are excited to share work that takes a difficult and important first step towards assessing clinician bias in real world data, and we hope this work helps others build more equitable approaches to bias detection.

% to be included in final version
\acks{We thank Ipek Ensari PhD, Ashwin Sawant MD PhD, and Matthew O’Connell PhD for their invaluable contributions to this research as members of the first author’s advisory committee. We also thank the sentiment annotators from previous projects who made this work possible. Special appreciation goes to the patients of Mount Sinai – may their data always be used for their benefit.}

\bibliography{CHIL_references}

\appendix
\section{}\label{apd:first}
See additional figures and tables below.

\onecolumn
\input{model_results_table}
% \twocolumn

\begin{figure*}[htbp]
 % Caption and label go in the first argument and the figure contents
 % go in the second argument
\floatconts
  {fig:nodes}
  {\caption{Marginal Effects of Patient Race/Ethnicity and Sex on SCZ Diagnosis at levels of SES.}}
  {\includegraphics[width=\linewidth]{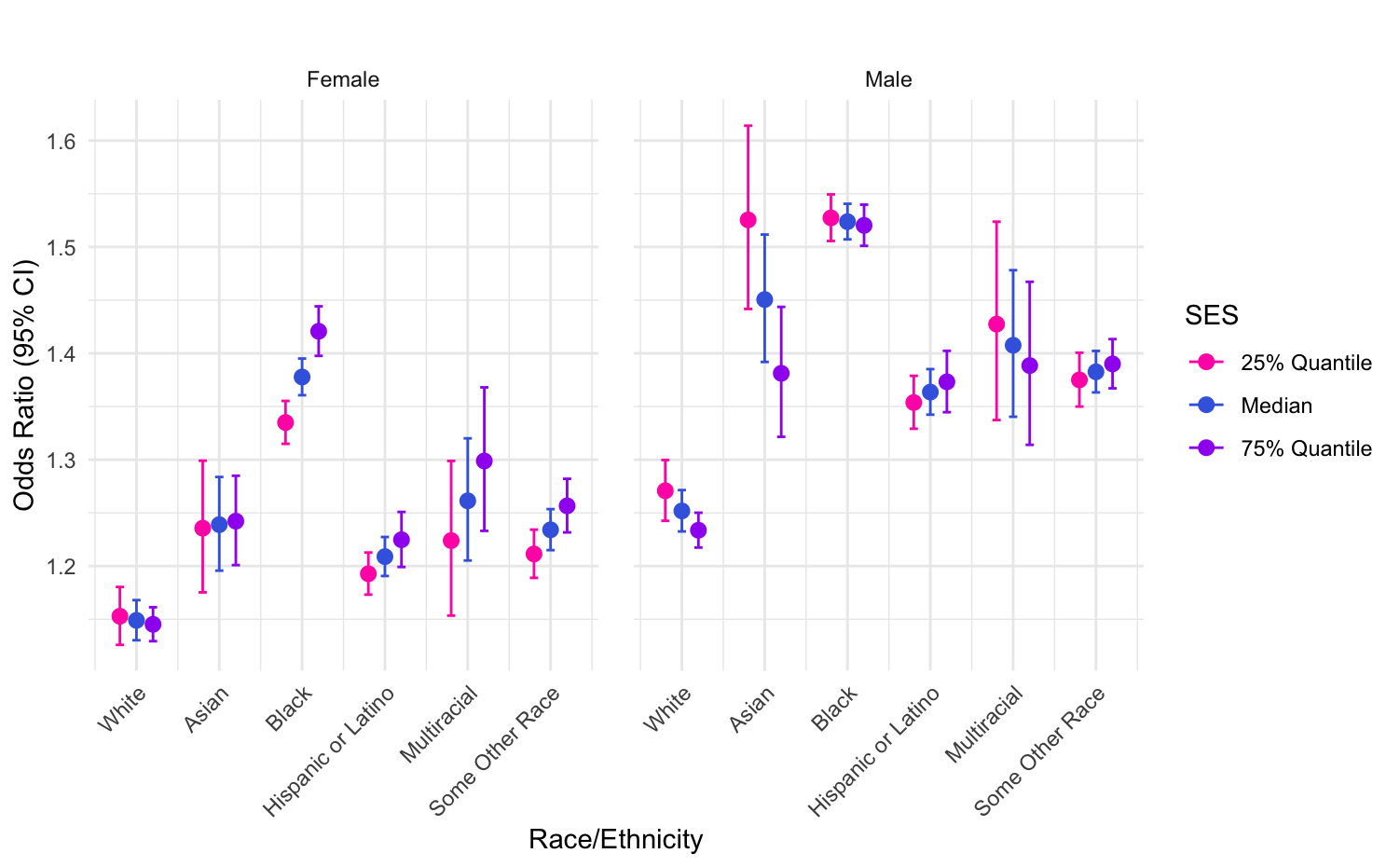}}
  \label{fig:ses_sex_race}
\end{figure*}

\begin{figure*}[htbp]
 % Caption and label go in the first argument and the figure contents
 % go in the second argument
\floatconts
  {fig:nodes}
  {\caption{Group Comparison of Number of Sentences across Race/Ethnicity}}
  {\includegraphics[width=\linewidth]{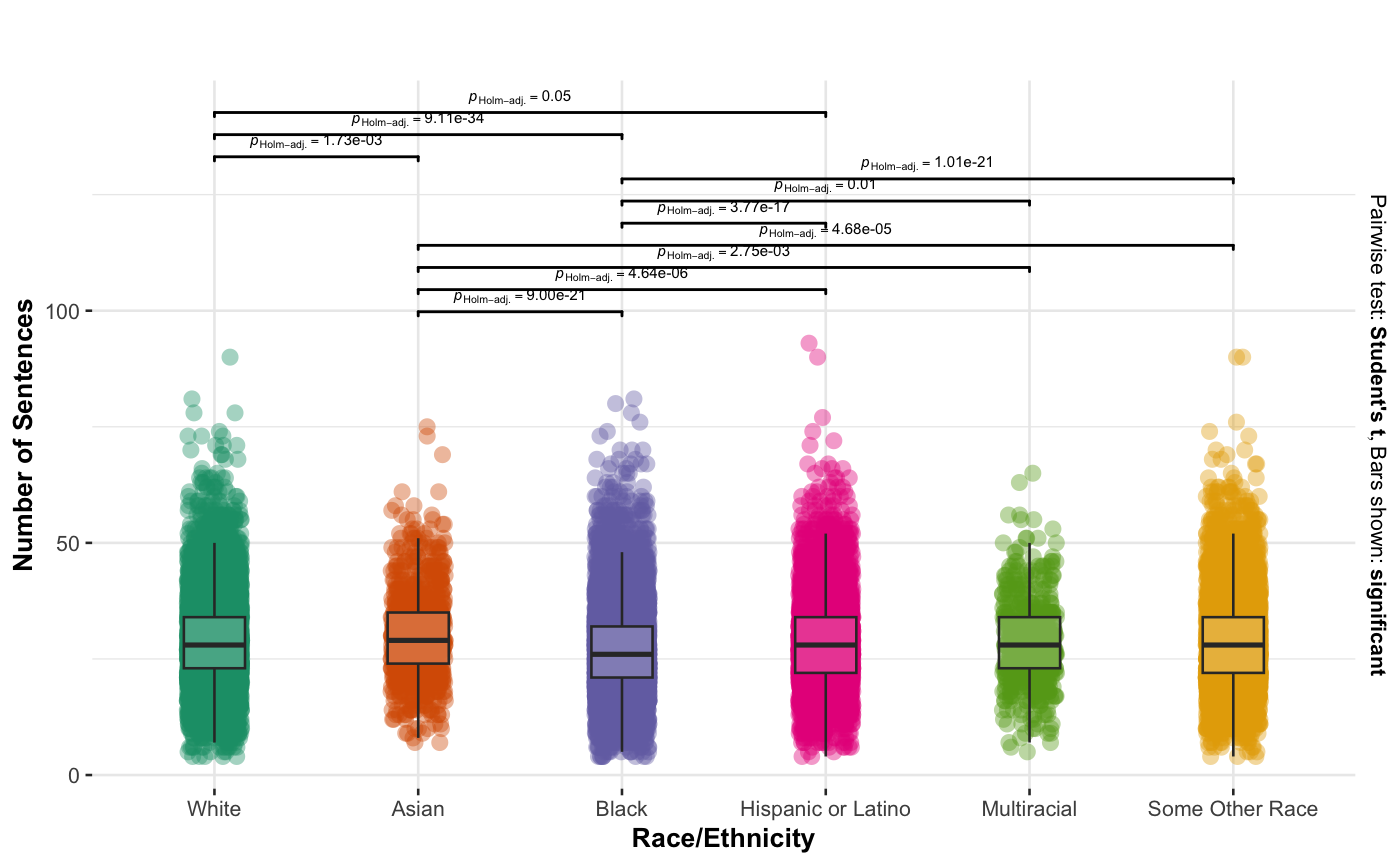}}
  \label{fig:group_n_sent}
\end{figure*}

\begin{figure*}[htbp]
 % Caption and label go in the first argument and the figure contents
 % go in the second argument
\floatconts
  {fig:nodes}
  {\caption{Group Comparison of Number of Negative Sentences across Race/Ethnicity}}
  {\includegraphics[width=\linewidth]{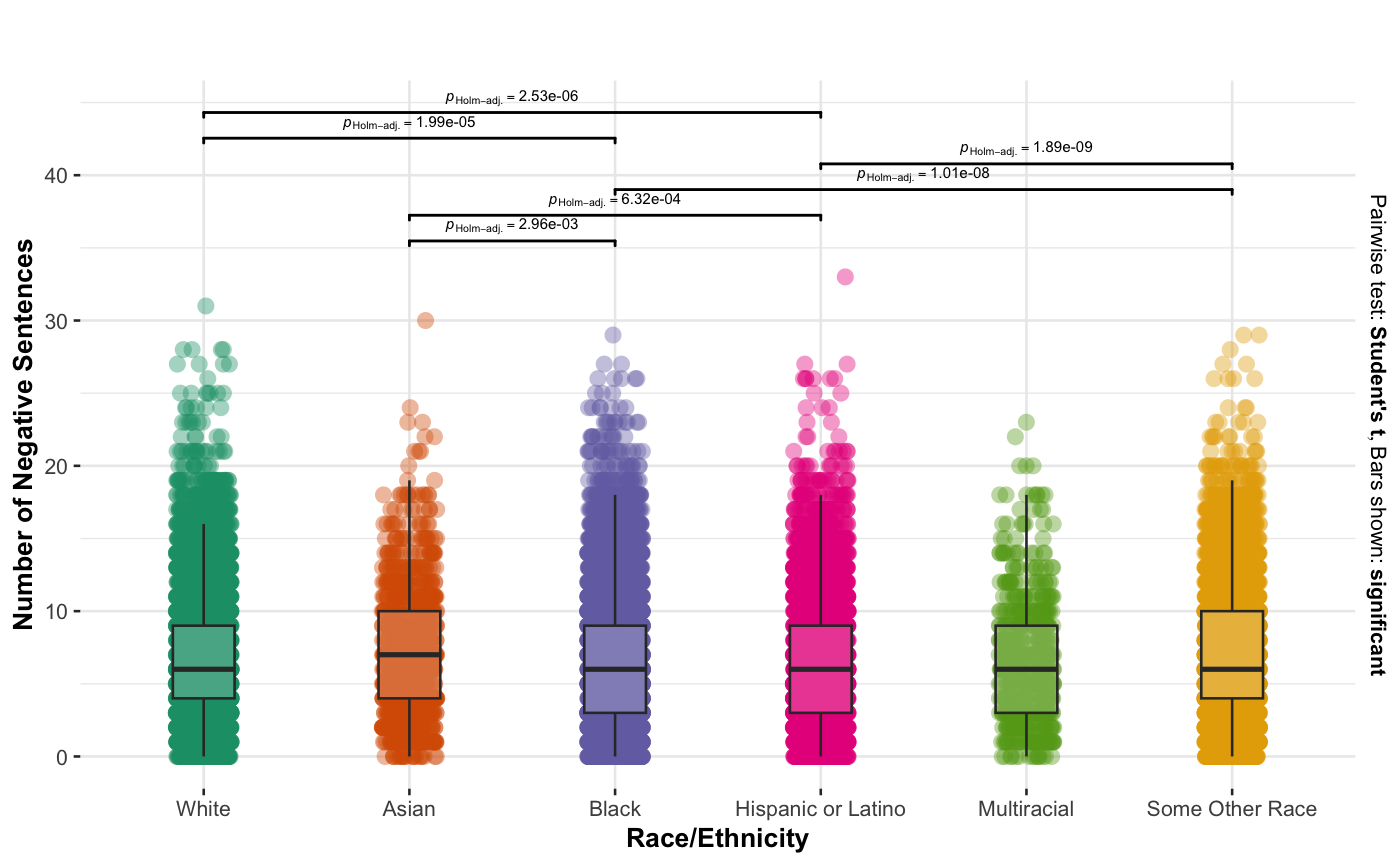}}
  \label{fig:group_neg_sent}
\end{figure*}

\begin{figure*}[htbp]
 % Caption and label go in the first argument and the figure contents
 % go in the second argument
\floatconts
  {fig:nodes}
  {\caption{Group Comparison of NSR across Race/Ethnicity}}
  {\includegraphics[width=\linewidth]{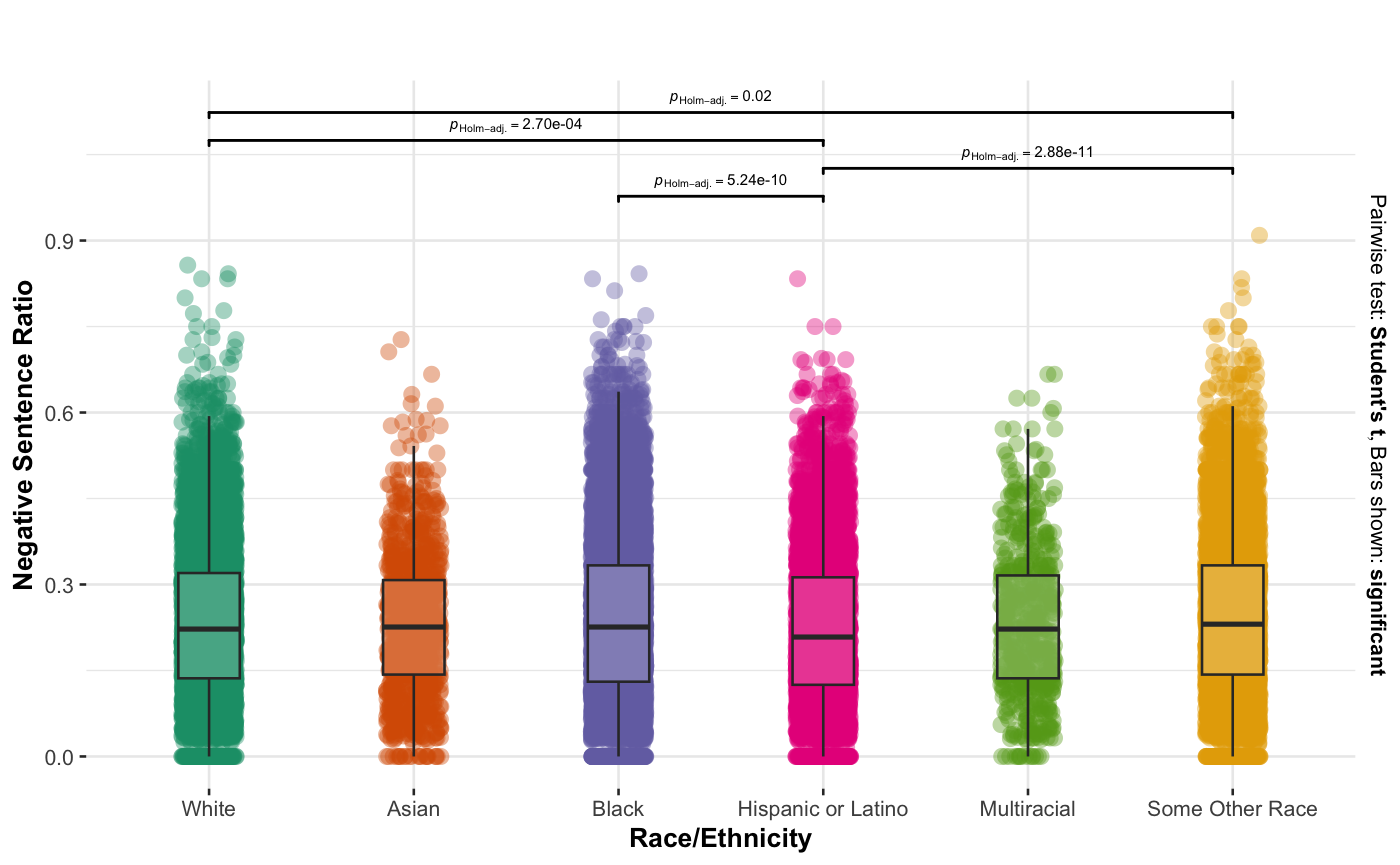}}
  \label{fig:group_nsr}
\end{figure*}

\end{document}

%% file: model_performance.tex
\begin{table}[hbtp]
\floatconts
  {tab: model-performance}
  {\caption{Manual Validation of Mistral Sentiment Labels}}
  {\begin{tabular}{lccc}
    \toprule
     & \textbf{Precision} & \textbf{Recall} & \textbf{F1}\\
    \midrule
    Negative & 0.78 & 0.70 & 0.74 \\
    Neutral & 0.52 & 0.87 & 0.65 \\
    Positive & 1.00 & 0.43 & 0.60 \\
    Macro Average & 0.77 & 0.67 & 0.66 \\
    \bottomrule
  \end{tabular}}
  \label{tab:llm_valid}
\end{table}

%% file: demo_table.tex
\begin{table*}[hbtp]
\floatconts
  {tab:schizophrenia-stratified}
  {\caption{Sample Characteristics Stratified by Schizophrenia%
  \protect\footnotemark}}
  {%
  \begin{tabular}{lccc}
    \toprule
     & \textbf{Non-SCZ Psych Patients} & \textbf{SCZ Patients} &  \\
    \textbf{Variable} & \textbf{n=20891} & \textbf{n=8114} & \textbf{p-value*} \\
    \midrule
    Sex = Male (count [\%]) & 10463 (50.1) & 4883 (60.2) & $\mathbf{<0.001}$ \\
    Race/Ethnicity (count [\%]) &  &  & $\mathbf{<0.001}$ \\
    \quad Asian & 701 (3.4) & 274 (3.4) & \\
    \quad Black & 5991 (28.7) & 3340 (41.2) & \\
    \quad Hispanic or Latino & 4221 (20.2) & 1284 (15.8) & \\
    \quad Multiracial & 419 (2.0) & 153 (1.9) & \\
    \quad Some Other Race & 4109 (19.7) & 1634 (20.1) & \\
    \quad White & 5450 (26.1) & 1429 (17.6) & \\
    Age (mean [SD]) & 36.3 (17.1) & 40.7 (15.4) & $\mathbf{<0.001}$ \\
    SES (mean [SD]) & 83272 (39755) & 81777 (38693) & $\mathbf{<0.005}$ \\
    History of Trauma-Related Disorder (count [\%]) & 11463 (54.9) & 1186 (14.6) & $\mathbf{<0.001}$ \\
    History of Substance Use Disorder (count [\%]) & 2513 (12.0) & 998 (12.3) & 0.539 \\
    Negative Sentences (mean [SD]) & 6.3 (4.2) & 7.8 (4.6) & $\mathbf{<0.001}$ \\
    Number of Sentences (mean (SD]) & 28.2 (9.3) & 28.4 (9.4) & 0.292 \\
    Negative Sentence Ratio (mean [SD]) & 0.22 (0.13) & 0.28 (0.14) & $\mathbf{<0.001}$ \\
    \bottomrule
  \end{tabular}
  }
\end{table*}
\footnotetext{*Chi-square test for categorical data and ANOVA for continuous data.}

%% file: diagnoses_or_table.tex
\begin{table}[hbtp]
\floatconts
  {tab:diagnosis-odds}
  {\caption{Odds of Anxiety, Bipolar, Depression, Trauma, and SCZ Diagnosis Given the NSR}}
  {\begin{tabular}{lc}
    \toprule
    \textbf{Diagnosis} & \textbf{OR (95\% CI)}\\
    \midrule
    Anxiety & 0.73 (0.75-0.71) \\
    Bipolar & 1.15 (1.18-1.12) \\
    Depression & 1.41 (1.46-1.36) \\
    Trauma & 0.51 (0.53-0.49) \\
    SCZ & 1.80 (1.87-1.74) \\
    \bottomrule
  \end{tabular}}
  \label{tab:diagnosis-odds}
\end{table}

%% file: model_results_table.tex
\begin{longtable}{lcccccc}
    \caption{Logistic Regression Results: Model With and Without NSR Terms} \\
        \toprule
        \textbf{Characteristic} & \multicolumn{3}{c}{\textbf{Model with NSR Terms}} & \multicolumn{3}{c}{\textbf{Model without NSR Terms}} \\
        \cmidrule(lr){2-4}\cmidrule(lr){5-7}
        & \textbf{OR} & \textbf{95\% CI} & \textbf{p-value} & \textbf{OR} & \textbf{95\% CI} & \textbf{p-value} \\
        \midrule
        \endfirsthead
        
        \multicolumn{7}{c}{\textbf{Table \thetable\ (continued)}}\\
        \toprule
        \textbf{Characteristic} & \multicolumn{3}{c}{\textbf{Model with NSR Terms}} & \multicolumn{3}{c}{\textbf{Model without NSR Terms}} \\
        \cmidrule(lr){2-4}\cmidrule(lr){5-7}
        & \textbf{OR} & \textbf{95\% CI} & \textbf{p-value} & \textbf{OR} & \textbf{95\% CI} & \textbf{p-value} \\
        \midrule
        \endhead
        
        \midrule
        \multicolumn{7}{r}{\small Continued on next page}\\
        \endfoot
        
        \bottomrule
        \multicolumn{7}{l}{\footnotesize Abbreviation: OR = Odds Ratio. CI = Confidence Interval.}\\
        \endlastfoot
        
        (Intercept) & 1.14 & 1.09, 1.20 & $\mathbf{<0.001}$ & 1.23 & 1.18, 1.28 & $\mathbf{<0.001}$ \\
        Substance & 0.96 & 0.94, 0.98 & $\mathbf{<0.001}$ & 0.97 & 0.95, 0.99 & $\mathbf{<0.01}$ \\
        Trauma & 0.71 & 0.70, 0.72 & $\mathbf{<0.001}$ & 0.69 & 0.69, 0.70 & $\mathbf{<0.001}$ \\
        Age & 1 & 1.00, 1.00 & $\mathbf{<0.001}$ & 1 & 1.00, 1.00 & $\mathbf{<0.001}$ \\
        
        Race/Ethnicity & & & & & & \\
        \hspace{1em}White & --- & --- &  & --- & --- &  \\
        \hspace{1em}Asian & 1.04 & 0.94, 1.16 & 0.46 & 1.06 & 0.96, 1.16 & 0.23 \\
        \hspace{1em}Black & 1.05 & 1.00, 1.11 & 0.054 & 1.10 & 1.05, 1.15 & $\mathbf{<0.001}$ \\
        \hspace{1em}Hispanic/Latino & 0.97 & 0.92, 1.03 & 0.30 & 1.00 & 0.95, 1.06 & 0.90 \\
        \hspace{1em}Multiracial & 1.03 & 0.91, 1.17 & 0.62 & 1.00 & 0.90, 1.12 & 0.94 \\
        \hspace{1em}Some Other Race & 0.97 & 0.92, 1.03 & 0.30 & 1.02 & 0.96, 1.07 & 0.55 \\
        
        Sex & & & & & & \\
        \hspace{1em}Female & --- & --- &  & --- & --- &  \\
        \hspace{1em}Male & 1.14 & 1.08, 1.20 & $\mathbf{<0.001}$ & 1.13 & 1.07, 1.19 & $\mathbf{<0.001}$ \\
        
        SES & 1 & 1.00, 1.00 & 0.96 & 1 & 1.00, 1.00 & 0.56 \\
        
        Substance * Trauma & 1.11 & 1.08, 1.14 & $\mathbf{<0.001}$ & 1.12 & 1.08, 1.15 & $\mathbf{<0.001}$ \\
        
        Race/Ethnicity * Sex & & & & & & \\
        \hspace{1em}Asian * Male & 1.20 & 1.04, 1.38 & $\mathbf{<0.05}$ & 1.20 & 1.04, 1.38 & $\mathbf{<0.05}$ \\
        \hspace{1em}Black * Male & 1.08 & 1.01, 1.15 & $\mathbf{<0.05}$ & 1.07 & 1.00, 1.14 & $\mathbf{<0.05}$ \\
        \hspace{1em}Hispanic/Latino * Male & 1.02 & 0.95, 1.10 & 0.59 & 1.01 & 0.94, 1.09 & 0.71 \\
        \hspace{1em}Multiracial * Male & 1.12 & 0.95, 1.32 & 0.18 & 1.12 & 0.95, 1.32 & 0.19 \\
        \hspace{1em}Some Other Race * Male & 1.03 & 0.96, 1.11 & 0.39 & 1.03 & 0.96, 1.11 & 0.41 \\
        
        Race/Ethnicity * SES & & & & & & \\
        \hspace{1em}Asian * SES & 1 & 1.00, 1.00 & 0.65 & 1 & 1.00, 1.00 & 0.64 \\
        \hspace{1em}Black * SES & 1 & 1.00, 1.00 & $\mathbf{<0.001}$ & 1 & 1.00, 1.00 & $\mathbf{<0.001}$ \\
        \hspace{1em}Hispanic/Latino * SES & 1 & 1.00, 1.00 & $\mathbf{<0.05}$ & 1 & 1.00, 1.00 & $\mathbf{<0.05}$ \\
        \hspace{1em}Multiracial * SES & 1 & 1.00, 1.00 & 0.055 & 1 & 1.00, 1.00 & 0.057 \\
        \hspace{1em}Some Other Race * SES & 1 & 1.00, 1.00 & $\mathbf{<0.01}$ & 1 & 1.00, 1.00 & $\mathbf{<0.01}$ \\
        
        Sex * SES & & & & & & \\
        \hspace{1em}Male * SES & 1 & 1.00, 1.00 & 0.11 & 1 & 1.00, 1.00 & 0.10 \\
        
        Race\_Ethnicity * Sex * SES & & & & & & \\
        \hspace{1em}Asian * Male * SES & 1 & 1.00, 1.00 & $\mathbf{<0.05}$ & 1 & 1.00, 1.00 & 0.055 \\
        \hspace{1em}Black * Male * SES & 1 & 1.00, 1.00 & $\mathbf{<0.05}$ & 1 & 1.00, 1.00 & $\mathbf{<0.05}$ \\
        \hspace{1em}Hispanic/Latino * Male * SES & 1 & 1.00, 1.00 & 0.63 & 1 & 1.00, 1.00 & 0.52 \\
        \hspace{1em}Multiracial * Male * SES & 1 & 1.00, 1.00 & 0.21 & 1 & 1.00, 1.00 & 0.22 \\
        \hspace{1em}Some Other Race * Male * SES & 1 & 1.00, 1.00 & 0.90 & 1 & 1.00, 1.00 & 0.89 \\

        NSR & 1.34 & 1.19, 1.52 & $\mathbf{<0.001}$ &  &  &  \\
        
        Race/Ethnicity * NSR & & & & & & \\
        \hspace{1em}Asian * NSR & 1.08 & 0.88, 1.34 & 0.46 &  &  &  \\
        \hspace{1em}Black * NSR & 1.16 & 1.06, 1.28 & $\mathbf{<0.005}$ &  &  &  \\
        \hspace{1em}Hispanic/Latino * NSR & 1.16 & 1.04, 1.30 & $\mathbf{<0.01}$ &  &  &  \\
        \hspace{1em}Multiracial * NSR & 0.88 & 0.68, 1.15 & 0.36 &  &  &  \\
        \hspace{1em}Some Other Race * NSR & 1.20 & 1.08, 1.34 & $\mathbf{<0.001}$ &  &  &  \\
        
        Sex * NSR & & & & & & \\
        \hspace{1em}Male * NSR & 0.95 & 0.89, 1.02 & 0.14 &  &  &  \\
        
        SES * NSR & 1 & 1.00, 1.00 & 0.33 &  &  &  \\
\end{longtable}
\label{tab:regression}
\footnotetext{Abbreviations: OR = Odds Ratio. CI = Confidence Interval.}